\documentclass[]{aastex631}

\usepackage{xcolor}
\usepackage{graphicx}
\usepackage{wrapfig}
\usepackage{float}

\begin{document}

\title{Witnessing a Galaxy Cluster Merger with JWST and \\a Chandra X-ray Temperature Map}

\author[0000-0003-0025-6762]{Brian C. Alden}
\affiliation{Center for Astrophysics and Space Astronomy \\
Department of Astrophysical and Planetary Sciences \\
University of Colorado Boulder \\
Boulder, CO 80309, USA}

\author[0000-0002-4468-2117]{Jack O. Burns}
\affiliation{Center for Astrophysics and Space Astronomy \\
Department of Astrophysical and Planetary Sciences \\
University of Colorado Boulder \\
Boulder, CO 80309, USA}

\begin{abstract}
    The first James Webb Space Telescope (JWST) image released was of galaxy cluster SMACSJ0723.3-7327, a lensing cluster at z=0.39 showing detail only JWST can provide. While the majority of the focus has been on the brilliantly lensed galaxies at redshifts far beyond it, there is more to the story than it being just a lensing cluster. The Chandra X-ray temperature map tells a tale of a merging cluster with a significant subcluster leaving a wake in the intracluster medium (ICM). This paper presents a high fidelity  temperature map of SMACSJ0723.3-7327 using adaptive circular binning, overlaid with the JWST image, showing clear signs of merger activity. As the ICM extends well past the boundaries of the JWST imagery, and no low-frequency radio observations are yet published, a fuller story of this cluster remains to be told. This new X-ray temperature map reveals new details of a moderately distant actively merging cluster.  
\end{abstract}


\section{Introduction} 
The hot intracluster medium (ICM) of the largest gravitationally bound objects in the universe, galaxy clusters, can tell tales not told by optical observation. At hundreds of millions of degrees Kelvin, the ICM emits X-ray radiation observed in this case by the Chandra X-ray Observatory (CXO). These images of the X-ray surface brightness show the true scale of the galaxy cluster by imaging the ICM, the hot gas between the constituent galaxies containing the vast majority of the baryonic matter present in the cluster and the universe. The first image released by NASA from the James Webb Space Telescope (JWST) was of a galaxy cluster exhibiting remarkable lensing properties, SMACSJ0723.3-7327 \citep{jwst-early-release}. While the released image shows a considerable amount of the cluster and its constituent galaxies, the X-ray surface brightness image shows the true extent of the cluster (see Figure 1). This X-ray image can be used to do spectral fitting in order to approximate the temperature across the ICM. This temperature paints a picture not seen by the X-ray surface brightness image alone or even when viewed with the JWST image. This combined data set reveals a galaxy cluster currently undergoing a merger.

\section{Data}

The X-ray Temperature map was generated from Chandra observation ID 15296, a 20 kilosecond observation. It was processed using `ClusterPyXT', a code that we developed to more easily generate temperature maps from calibrated X-ray data. This temperature map is the product of $\sim$6000 spectral fits across the image using adaptive circular binning. This is a process where a circular bin is chosen for each pixel in the image with the bin radius varying to achieve the desired signal to noise (50), with a maximum size of 100 pixels. While some spatial resolution is lost with this process, the end result is a high-quality temperature map that provides a qualitative view of the cluster and can guide further study. A full description of this process is found in \cite{clusterpyxt}. The ClusterPyXT code also generates a temperature error map which in this case is about $ 10-25\%$. Furthermore, we generated 1-D profiles in wedges with annular bins to verify the temperatures seen in Figure 1.

The JWST NIRCam image of the cluster was constructed using the process outlined by the Space Telescope Sciences Institute\footnote{\url{https://jwst-docs.stsci.edu/getting-started-with-jwst-data}}. JWST made six observations with NIRCam of SMACSJ0723.3-7327, one for each filter, ranging from $\sim$15 to 60 kiloseconds. An RGB FITS image was made from three of the six observations, allowing for alignment with the X-ray surface brightness and temperature map via WCS coordinates. 

Once aligned, these images were combined in Adobe Photoshop in order to see the temperature, X-ray surface brightness contours, and the JWST observation. The NASA published image\footnote{\url{https://webbtelescope.org/contents/media/images/2022/038/01G7JGTH21B5GN9VCYAHBXKSD1}} \citep{jwst-early-release}, which lacks coordinate information, captures all 6 observations and produces a sharper image. The opacity and blending modes were selected to produce the combination which attempts to show the best of each image. While there are certainly biases introduced by this process, the resulting image still provides a level of insight not garnered with the standalone images.

\section{Discussion}
The temperature map for SMACSJ0723.3-7237, generated by ClusterPyXT\footnote{\url{https://github.com/bcalden/ClusterPyXT}}, illustrates that despite looking semi-relaxed in the X-ray surface brightness (represented by the white contour lines in Figure 1), it is far from so. The cooler region towards the top center of the figure along with the heated region in the bottom left show a cluster that is far from equilibrium. This temperature difference is indicative of a cluster undergoing a merger. These temperature measurements agree with the weighted Voronoi tessellation (WVT) binned temperature map in the upcoming work of \cite{jwst-xray}. Further evidence pointing to a cluster merger comes from redshift measurements of some constituent galaxies from zMUSE showing that the distribution is non-Gaussian \citep{jwst-modeling}. A relaxed cluster would have a Gaussian distribution of redshifts. While there are not enough redshift measurements for the redshifts alone to be suggestive of a merger as a sampling bias may be at play, viewed with the temperature map they both suggest we are witness to a merger. 

With the combined image, it appears that some of the centrally bright galaxies in the cluster are in the process of merging with a second cluster. There is a small wake in the X-ray surface brightness coinciding with this infalling subcluster, the cooler region in the temperature map shows this wake to a greater degree. The coolest part of the colder region near the center of the cluster measures $5.3 \pm 0.6$ keV.  In the heated region opposite of it, the peak measures $10.0 \pm 2.1$ keV, possibly indicative of a shock. These two temperature features strongly align and suggest a mildly supersonic infalling subcluster of appreciable size. Unfortunately, there is insufficient counts in the current X-ray data to verify a significant pressure jump that would validate a shock.  Further X-ray integration with Chandra is necessary to generate enough data for a statistically significant claim. With that said, the picture this temperature map presents suggests it is a strong candidate for a shock in the periphery of the cluster. 

SMACSJ0723.3-7237 appears to be a merging cluster, and as such, new radio observations taken by MeerKAT recently may show a radio relic in this general region or just beyond the heating region/shock. While it is difficult to predict precisely where it may be due to the projection effects and limited kinematic data, the direction suggested by the temperature map would make for good hunting grounds with low-frequency radio observations \citep{govoni-radio,a115-hallman,2010-vanweeren,2019-vanweeren}. 

Much of the discussion to date on SMACSJ0723.3-7237 is around the lensed galaxies, but there is more going on in than just gravitational lensing. The first image released by JWST is also that of a merging cluster which is more dynamic than at first glance. 

\begin{figure}
    \centering
    \centerline{\includegraphics[width=18cm]{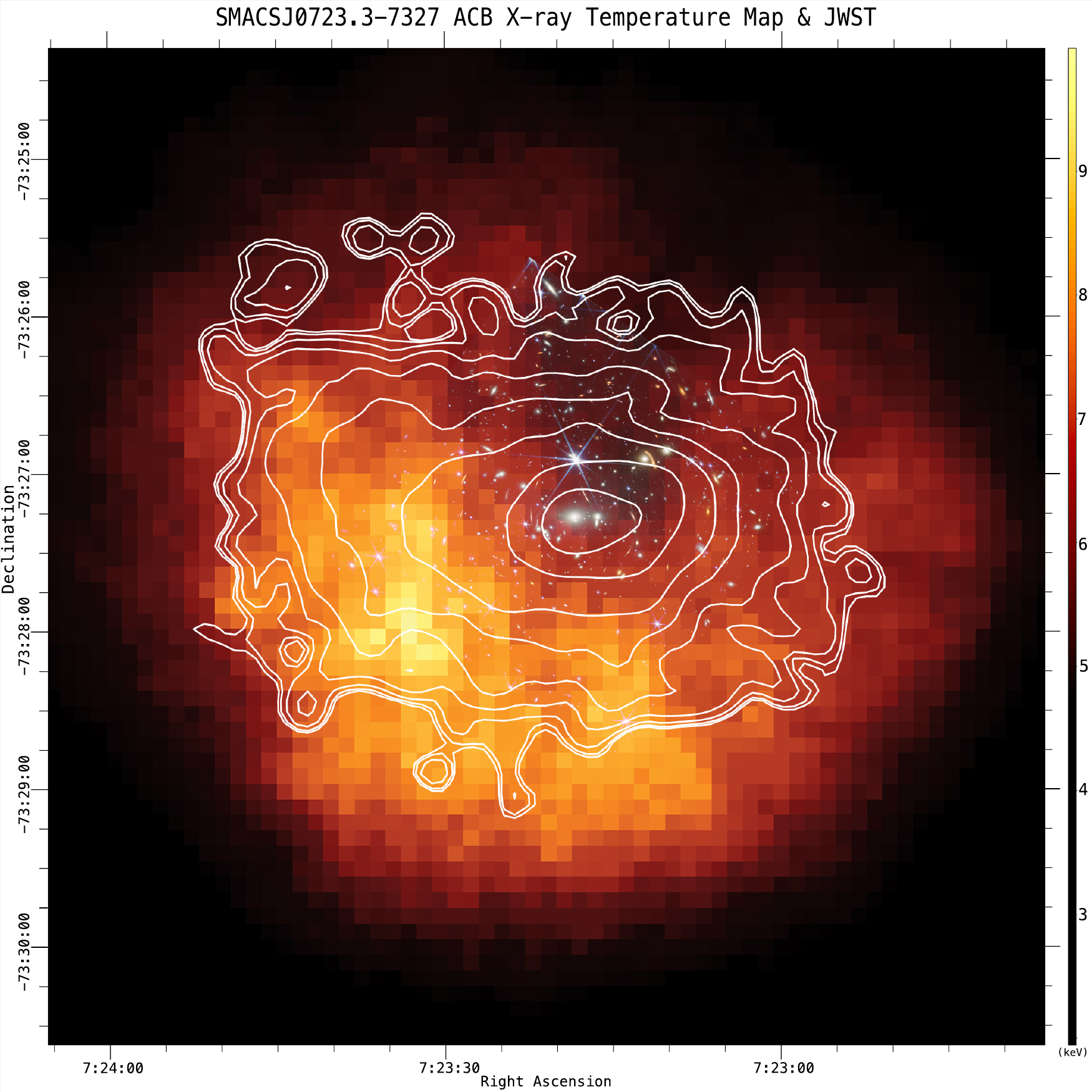}}
    \caption{Combined ACB X-ray temperature map and JWST infrared image across 6 bands. White contour lines represent the X-ray surface brightness from Chandra X-ray Observatory. The logarithmic contour lines shown begin at 3$\times$ the RMS background of the image and ascend to just below the peak surface brightness. The cooler region in the north just right of center provides evidence of an infalling subcluster. The coolest portion near the center of the cluster is approximately $5.3 \pm 0.6$ keV.  The heated region opposite of it alludes to a shock front from this merger and has a peak temperature of $10.0 \pm 2.1$ keV. While more data is necessary to make the claim that a shock is present, the current evidence points in that direction.}
    \label{fig:JWST-Temp-XSB}
\end{figure}

\clearpage
\bibliography{main}{}
\bibliographystyle{aasjournal}
\end{document}